\documentclass[a4paper,fleqn,usenatbib]{mnras}

\usepackage[T1]{fontenc}
\usepackage{ae,aecompl}

\usepackage{graphicx}	
\usepackage{amsmath}	
\usepackage{amssymb}	

\def\kms{{\rm\,km\,s^{-1}}}

\def\pc{{\rm\,pc}}
\def\kpc{{\rm\,kpc}}
\def\myr{{\rm\,Myr}}

\def\masyr{{\rm\,mas \, yr^{-1}}}

\def\gtsima{$\; \buildrel > \over \sim \;$}
\def\simgt{\lower.5ex\hbox{\gtsima}}

\title[Collision between two open clusters]{First evidence of a collision between two unrelated open clusters in the Milky Way}

\author[A.E. Piatti and K. Malhan]{
Andr\'es E. Piatti$^{1,2}$\thanks{E-mail: andres.piatti@unc.edu.ar} and Khyati Malhan$^{3,4}$\\
$^{1}$Instituto Interdisciplinario de Ciencias B\'asicas (ICB), CONICET-UNCUYO, Padre J. Contreras 1300, M5502JMA, Mendoza, Argentina\\
$^{2}$Consejo Nacional de Investigaciones Cient\'{\i}ficas y T\'ecnicas, Godoy Cruz 2290, C1425FQB,  Buenos Aires, Argentina\\
$^{3}$The Oskar Klein Centre for Cosmoparticle Physics, Department of Physics, Stockholm 
University, AlbaNova, 10691 Stockholm, Sweden\\
$^{4}$Max-Planck-Institut f\"ur Astronomie, K\"onigstuhl 17, D-69117, Heidelberg, Germany
}

\date{Accepted XXX. Received YYY; in original form ZZZ}

\pubyear{2021}

\begin{document}
\label{firstpage}
\pagerange{\pageref{firstpage}--\pageref{lastpage}}
\maketitle

\begin{abstract}
We report the first evidence of an on-going collision between two star clusters in our Galaxy, namely IC~4665 and Collinder~350. These are open clusters located at a distance of $\sim330$~pc from the Sun and $\sim$100~pc above the Galactic plane, and they both have prograde motions with only a small difference in their velocities (Collinder~350 moves $\sim5\,\rm{km\,s^{-1}}$ faster than IC~4665); as inferred from ESA/{\it Gaia} based catalogue. Interestingly, the two clusters are physically separated by only $\sim$36~pc in space; a distance that is smaller than the sum of their respective radii. Furthermore, the clusters exhibit signatures of elongated stellar density distributions, and we also detect an onset of an inter-cluster stellar bridge. Moreover, the orbit analysis suggests that the younger cluster IC~4665 (age=53 Myr) must have formed at a distance $>500$~pc away from Collinder~350 (age=617 Myr). These findings together imply that the two clusters do not represent merging of two objects in a binary system, rather, what we are witnessing is an actual collision between two independently formed star clusters. This collision phenomenon provides a unique opportunity to explore new aspects of 
formation and evolution theory of star clusters.
\end{abstract} 




\section{Introduction}

The Milky Way is populated by thousands of open clusters, and most of them observed as independent objects. Here, we use the classification of open cluster given by \citet[][see their Table~1]{bicaetal2019}, who differentiated 
open clusters from associations, embedded clusters, cluster remnants, etc 
\citep[see, e.g.][]{feigelsonetal2011,kounkeletal2018,townsleyetal2018,cantatgauditetal2019b,keretal2021}.
Binary clusters are pairs of clusters that posses small physical separation.

\citet{dlfm2010} estimated that nearly $10\%$ of the total population of open clusters could in fact be binary systems. However, the number of physical pairs identified by \citet{dlfm2009} increases up to only 34 candidates, with a best candidate sample of only 19 bonafide pairs.  Interestingly,
\citet{dlfm2009}  arrived at this result based on the distinction between couple and binary clusters,
after considering that clusters may form as pairs, triplets or higher multiplicity star clusters.
\citet{conradetal2017} used radial velocities from the RAdial Velocity Experiment \citep[RAVE][]{steinmetzetal2006} catalogue, and also other catalogues, to identify only 19 pairs, and among these pairs only 14 pairs possessed physical 
 separations smaller than $100\pc$. \citet{soubiranetal2018} employed the {\it Gaia} DR2 database  \citep{gaiaetal2016, gaiaetal2018b} to analyse 861 open clusters in the local 6D phase space volume and identified only 5 physical pairs. They concluded that all the previously identified binary clusters were chance alignments.  It is important to find bonafide pairs of open clusters as they provide ideal laboratories to constrain the still poorly understood formation and evolution theory of star clusters.

The above overview reveals that binary open clusters are not very common in the Milky Way. Indeed, \citet{dlfm2010} showed from simulations that long-lasting binary open clusters are not very stable systems. They found that cluster pairs that initially form together quickly undergo merging. They reported that the physical separation between the pairs peaks at $25-30\pc$.
With the aim of examining the above results, here we used the recently updated open cluster catalogue compiled by \citet{diasetal2021}. This catalogue includes homogeneous astrophysical properties derived for 1743  open clusters from {\it Gaia} DR2 data  sets 
\citep{gaiaetal2016,gaiaetal2018b}. We performed a search for cluster pairs with physical separation smaller than $40\pc$ and
compared those distances with the sum of their adopted radii ($r_1+r_2$). To calculate the corresponding uncertainties we used a Monte Carlo sampling of the involved parameters and their uncertainties. 

\begin{figure}
\includegraphics[width=\columnwidth]{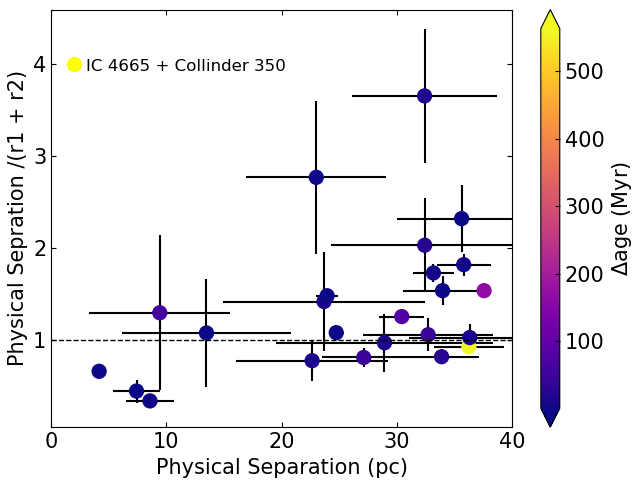}
\caption{Ratio of the physical separation to the sum of the cluster  adopted radii
versus 3D cluster separation. Cluster pairs below the dotted line are thought to be
experiencing merging.}
\label{fig1}
\end{figure}

Figure~\ref{fig1} shows that out of 1743 open clusters, the number of close pairs 
with separations smaller than 40 pc is relatively small (although we note that at this stage 
we did not consider the relative velocities between the two clusters).
 All detected pairs, but one, are composed of open clusters that possess similar ages ($\Delta$age $\la$ $100\myr$), implying that these binary pairs were formed at the same time (likely at the same location in our Galaxy). This inference is in very good agreement with our previous knowledge of the age difference of physical binary clusters \citep{dlfm2009, conradetal2017, soubiranetal2018}. We additionally note that those binary clusters with ratios 
 of the physical separation to the sum of the tidal radii  smaller than the unity (dotted line in Fig.~\ref{fig1}) are experiencing merging  \citep{dlfm2010}. Interestingly, Figure~\ref{fig1} 
 shows that there exists one cluster pair, located
 below the dotted line, with an age difference $>500\myr$ . This peculiar cluster pair corresponds to the system comprising of two open clusters -- IC~4665 and Collinder~350.

In this  work, our aim is to report the discovery of this first observed collision between these two open clusters IC~4665 and Collinder~350. In Section~2, we describe the observation evidence that supports the cluster collision, which Section~3 deals with numerical simulations of the clusters' orbits. Section~4 presents the final conclusions and discussion of this work.

\begin{figure}
\includegraphics[width=\columnwidth]{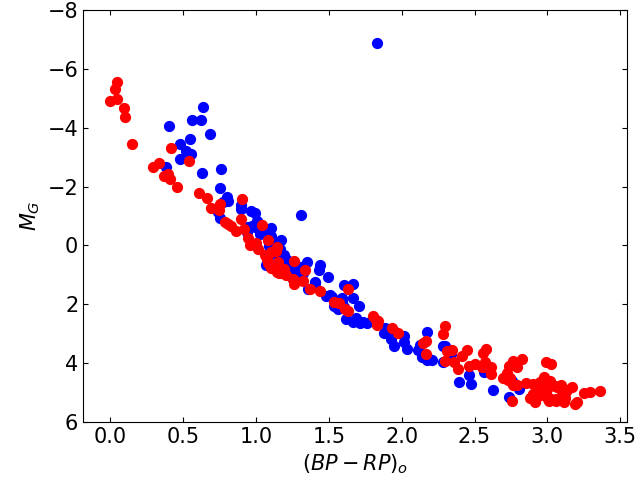}
\caption{Colour-magnitude diagram for cluster members in Collinder~350 and IC~4665, represented
with filled blue and red circles, respectively.}
\label{fig2}
\end{figure}
\section{The observed cluster collision}

IC~4665 has an age of $53 \myr$, and its phase-space position in the Galaxy in heliocentric coordinates is $(\rm{ra, dec., d_{\odot}, \mu^{*}_{\alpha}, \mu_{\delta}, v_{r}})$ = ($266.5541\degr$, $5.5800\degr$, $0.329 \pm 0.015\kpc$, $-0.896 \pm 0.317 \masyr$, $-8.504 \pm 0.328 \masyr$, $-14.142 \pm 4.351 \kms$). Collinder~350 has an age of 
$617 \myr$, and its phase-space position is $(\rm{ra, dec., d_{\odot}, \mu^{*}_{\alpha}, \mu_{\delta}, v_{r}})$ = ($266.9397\degr$, $1.5473 \degr$, $0.356 \pm 0.010 \kpc$, $-4.928 \pm 0.412\masyr$, $-0.020 \pm 0.274 \masyr$, $-14.665 \pm 1.309 \kms$)
\citep{diasetal2021}.
As one can note, although IC~4665 and Collinder~350 are separated in the sky by $\sim 4\degr$, their volumes seem to be superimposed. This can be inferred when considering their dimensions as estimated by  \citet{diasetal2021}, i.e.,  their radii
estimated as the distance to the cluster centres of the farthest 6D phase space member star. We note that the  criterion to define the clusters' extensions (their radii) does
not invalidate the discovery of the collision between IC~4665 and Collinder~350 which,
at some level, is already indicated from Figure~\ref{fig1}. 

To analyze Collinder~350 and IC~4665 in more detail, we compiled lists of their member stars.
For both the clusters, we adopted those stars which possess probability
$P$=1 in \citet{diasetal2021}, which were recalculated
using all the stars with membership probabilities $P$=1 in \citet{cga2020}. To redetermine
membership probabilities \citet{diasetal2021} applied a variation of the classical maximum
likelihood approach described in \citet{diasetal2014b} and \citet{monteiroetal2020}. Membership
probabilities in \citet{cga2020} were determined by employing the 
unsupervised UPMASK code \citep{kmm2014}. Fig.~\ref{fig2} depicts the  colour-magnitude
diagram built using {\it Gaia} DR2 $G$,$BP$,$RP$ photometry for these cluster members. 
For comparison purposes
we corrected the magnitudes and colours by reddening using the mean cluster $E(B-V)$ values
 given by \citet{diasetal2021}, the heliocentric cluster distances mentioned above, and the
$A_\lambda / A_V$ coefficiente given by \citet{cetal89} and \citet{wch2019}. The fainter
Main Sequence turnoff, the more pronounced curvature of the upper Main Sequence and the
present of a red giant in Collinder~350 reveals its older age compared to IC~4665.

\begin{figure*}
\includegraphics[width=\columnwidth]{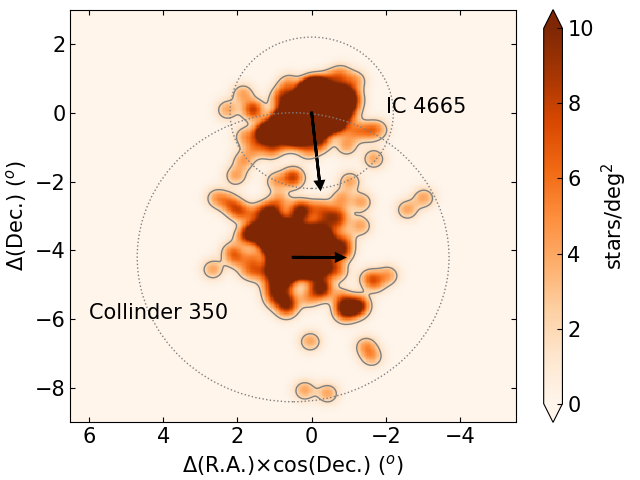}
\includegraphics[width=\columnwidth]{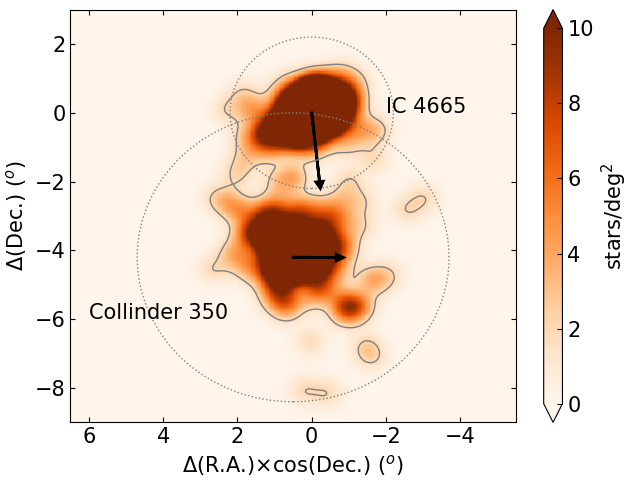}
\includegraphics[width=\columnwidth]{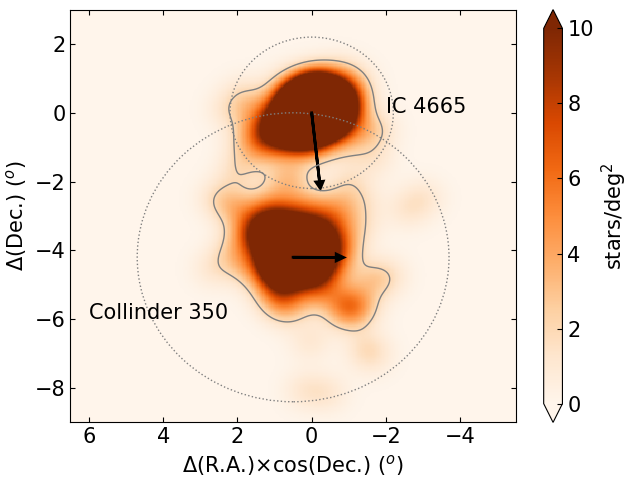}
\includegraphics[width=\columnwidth]{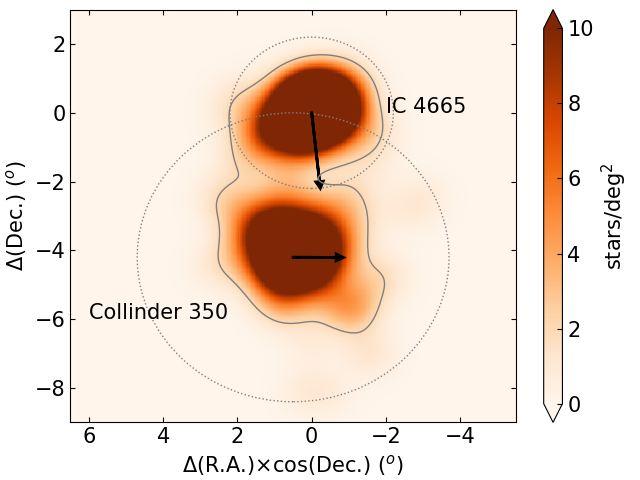}
\caption{Stellar density maps of the  partially merged cluster pair IC~4665 (northern cluster) and Collinder~350. Dotted circles and black arrows represent the cluster adopted radii and proper motion vectors, respectively. A gray-coloured contour for a stellar density of  2 stars/deg$^2$ is also drawn. The maps were generated with KDE bandwidths of 0.2, 0.3, 0.4, and 0.5 (from
top to bottom, and from left to right), respectively.}
\label{fig3}
\end{figure*}
\begin{figure}
\includegraphics[width=\columnwidth]{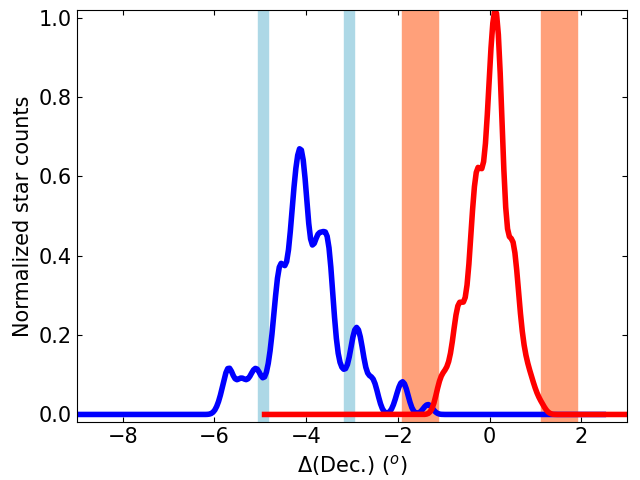}
\caption{Projected stellar density map  along the Declination axis. IC~4665 and
Collinder~350 are drawn with red and blue lines respectively. The vertical shaded
regions represent the tidal radii and their uncertainties as estimated by
\citet{piskunoveal2008}. Stars of both clusters have escaped their respective
tidal volumes.}
\label{fig4}
\end{figure}

Bearing in mind that both clusters are physically separated by a distance smaller than
the sum of  their radii (see Fig.~\ref{fig1}), their projected stellar structures should
show evidence of such a closeness. We note that the stellar density profiles of
tidally unperturbed clusters usually follow a nearly spherical \citet{king62} profile. Extended
stellar envelopes/haloes, scattered stellar debris, tidal tails, etc, are instead found around 
clusters that experience tidal interactions \citep[e.g.,][]{pangetal2021}.

Figure~\ref{fig3} shows the stellar density map of IC~4665 and Collinder~350. We build this map by using all cluster members in IC~4665 and  Collinder~350, the \texttt{scikit-learn} software machine learning library \citep{scikit-learn} and its kernel density estimator (KDE). We employed a grid of 500$\times$500 
boxes on the added cluster fields and allowed the bandwidth to vary from 0.1 up 0.5 in
steps of 0.1.
Fig.~\ref{fig3} shows that the projected cluster circles, delineated by their tidal radii,
intersect.  We counted the number of IC~4665's members embraced within the Collinder~350's radius 
and found that $\sim$ 58 percent of them are inside the later.
IC~4665, located to the north in this map, exhibits an elongated shape projected on the sky, while the outer region Collinder~350 
seems to be azimuthally non uniform and presents some stellar debris.  Both these features
are commonly found in clusters experiencing tidal interaction  \citep[e.g.,][]{dalessandroetal15}.
Figure~\ref{fig4} shows that the projected
stellar densities along the Declination axis  reveal that stars of both clusters
have crossed their respective tidal radii  estimated by \citet{piskunoveal2008}. Indeed, when
larger KDE bandwidths are used to build Fig.~\ref{fig3}, so that large spatial scale structures can be
highlighted, the line connecting both clusters (the gravitational bond line) along the
Declination axis is apparent.

\begin{figure}
\hbox{
\includegraphics[width=\columnwidth]{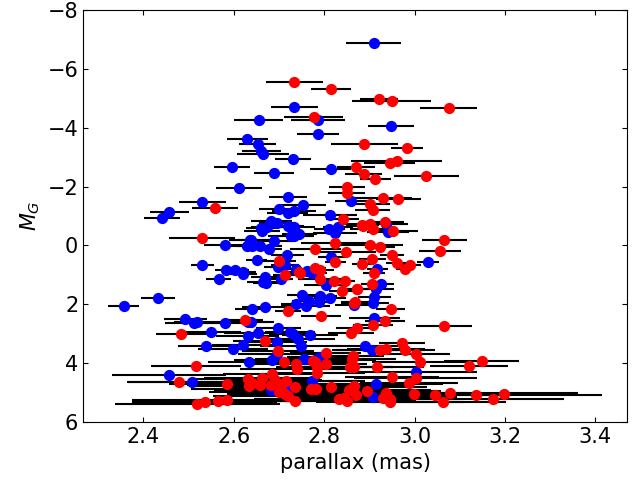}
}
\caption{Relation between the absolute magnitudes and the parallaxes of stars drawn
in Fig.~\ref{fig2}. Blue and red filled cirles correspond to Collinder~350 and IC~4665,
respectively. Individual error bars are also included. The data were kindly provided by
W. Dias.}
\label{fig5}
\end{figure}

The analysis of stellar parallaxes also confirms the proximity of both clusters along
the line-of-sight.
To this end we used the  \citet{diasetal2021}'s parallaxes of every star with membership probabillity 
$P$ =1, according to \citet{diasetal2021} and \citet{cga2020}. Both studies applied
appropriate filters according to \citet{gaiaetal2018c} to guarantee that only
stars with good astrometric solutions are used. Stars resulted in cluster members from the
likelihood approaches \citep{diasetal2021} and the $k$-means clustering technique \citep{cga2020} 
have parallaxes and proper motions that belong to a concentrated group of values to the light
of the uncertainties in proper motion, in parallax, and the correlation coefficients given
in the {\it Gaia} DR2 catalogue. Fig.~\ref{fig5} shows the {\it Gaia} DR2 parallax distribution for cluster
members in Collinder~350 and IC~4665, drawn with blue and red filled circles, respectively.
We  next determined the mean and dispersion of each cluster's parallax by
employing a maximum likelihood approach, similar to the method detailed in \citet{franketal2015}.
The relevance lies in accounting for each individual star's measurement errors, which could 
artificially inflate the dispersion if ignored. 
We optimized the probability $\mathcal{L}$ that a
given ensemble of stars with 
parallaxes $\varpi$$_i$ and parallax errors $\sigma_i$ are drawn from a population with mean
parallax $<$$\varpi$$>$ and intrinsic dispersion W  
\citep[e.g.,][]{Walker2006,Koch2018Gaia}, 
as follows:
\begin{eqnarray*}
\mathcal{L}\,=\,&\prod_{i=1}^N&\, \left(\,2\pi\left[\sigma_i^2 + W^2 \, \right]\,\right)^{-\frac{1}{2}} \\
&\times&\,\exp \left(-\frac{\left(\varpi_i \,- <\varpi>\right)^2}{\sigma_i^2 + W^2} \right), 
\end{eqnarray*}
where the errors on the mean and dispersion were computed from the respective covariance matrices. 
 We note that this approach assumes that the error distribution is Gaussian, which is adopted here 
because of the limited number of stars \citep[cf.][]{franketal2015}.

Table~\ref{tab:table1} shows
the mean and dispersion for both clusters for different magnitude ranges (see Fig.~\ref{fig5}).
As can be seen, the parallax errors do not play a role; except possibly for the faintest stars,
albeit the different number of stars in both clusters. We then defined a merger index as follows:

\begin{equation} 
\frac{(<\varpi> + W)_{\rm Collinder~350} - (<\varpi> - W)_{IC~4665}}{W_{Collinder~350}+W_{IC~4665}},
\end{equation}

\noindent which measures in percentage the fraction of intersection between both clusters' volumes. 
The resulting merger indices show that both clusters are partially merged.

\begin{table*}
\caption{Parallax mean and dispersion values for Collinder~350 and IC~4665.}
\label{tab:table1}
\begin{tabular}{@{}cccccccc}\hline
 & \multicolumn{3}{c}{Collinder~350} & \multicolumn{3}{c}{IC~4665} & \\
$M_G$ (mag) & $\varpi$ (mas) & W (mas)  & N & $\varpi$ (mas) & W (mas) & N & merger index ($\%$)\\\hline
$<$ -2.0    & 2.72$\pm$0.03 & 0.09$\pm$0.02 & 14 & 2.90$\pm$0.01 & 0.07$\pm$0.01 & 14 & --\\
-2.0 - 0.0  & 2.71$\pm$0.01 & 0.11$\pm$0.01 & 30 & 2.87$\pm$0.03 & 0.11$\pm$0.03 & 18 & 27\\
0.0 - 2.0   & 2.73$\pm$0.03 & 0.12$\pm$0.02 & 35 & 2.86$\pm$0.02 & 0.08$\pm$0.01 & 27 & 35\\
2.0 - 4.0   & 2.67$\pm$0.01 & 0.11$\pm$0.01 & 33 & 2.85$\pm$0.03 & 0.13$\pm$0.03 & 25 & 27\\
$>$ 4.0     & 2.74$\pm$0.07 & 0.11$\pm$0.60 & 8 & 2.80$\pm$0.05 & 0.10$\pm$0.05 & 55  & 68\\
\hline
\end{tabular}
\end{table*}

We took advantage of the proper motions, radial velocities and  parallaxes of individual cluster members  kindly provided by \citet{diasetal2021} to compute their Galactic positions and space velocity components. We computed Galactic coordinates $(X,Y,Z)$ and space velocities 
$(V_X,V_Y,V_Z)$ employing the \texttt{astropy}\footnote{https://www.astropy.org} package \citep{astropy2013,astropy2018}, which simply required the input of Right Ascension, Declination, 
 parallaxes, proper motions and radial velocity of each star. We adopted the default
values for the Galactocentric coordinate frame, namely : ICRS coordinates (RA, DEC) of the
Galactic centre = (266.4051$\degr$, -28.936175$\degr$); Galactocentric distance of the Sun =
8.122 kpc, height of the Sun above the Galactic midplane = 20.8 pc; and solar motion relative to the
Galactic centre = (12.9, 245.6, 7.78) km/s. The position of the Sun is assumed to be on the $X$ axis
of the final, right-handed system. That is, the $X$ axis points from the position of the Sun
projected to the Galactic midplane to the Galactic centre - roughly towards (l,b)=
(0$\degr$,0$\degr$). The $Y$ axis points roughly towards Galactic longitude {\it l}=90$\degr$, 
and the $Z$ axis points roughly towards the North Galactic Pole (b=90$\degr$).

\begin{figure*}
\includegraphics[width=\columnwidth]{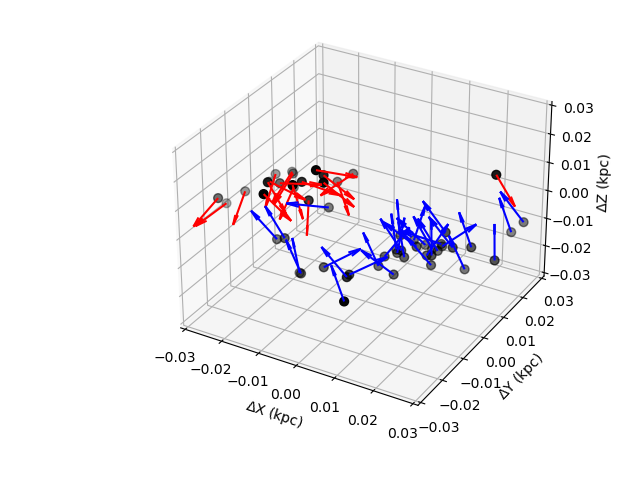}
\includegraphics[width=\columnwidth]{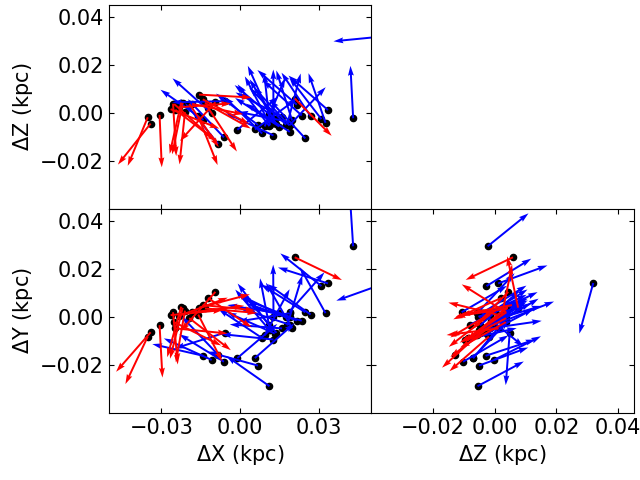}
\includegraphics[width=\columnwidth]{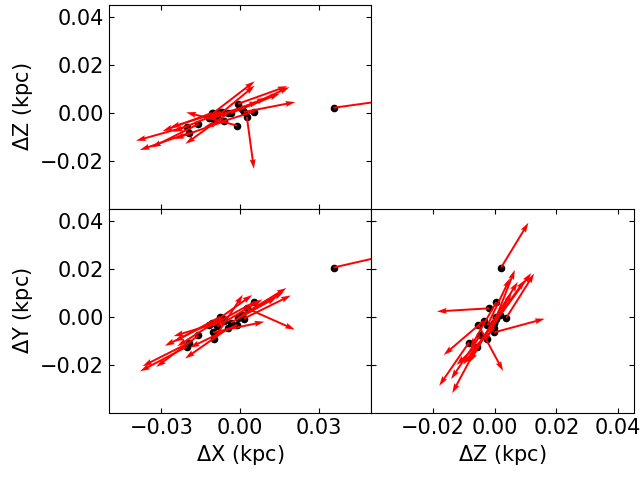}
\includegraphics[width=\columnwidth]{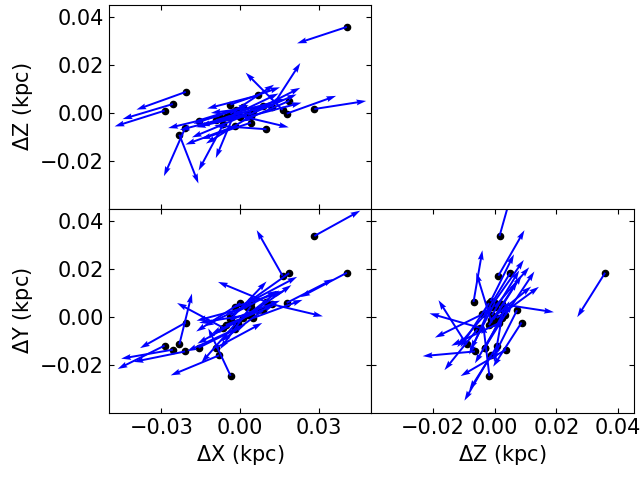}
\caption{{\it Left-top panel:} 3D view of IC~4665 and Collinder~350 members (black dots) with their space velocity vectors relative to the motion of the centre of mass (assumed to be the geometrical middle point between both clusters), drawn with red and blue arrows, respectively. 
{\it Right-top panel:} projected space velocity vectors of IC~4665 and Collinder~350, with respect to the motion of their centre of mass.  The Sun is placed at $\Delta$$X$=7.82 kpc,$\Delta$$Y$=
-0.17 kpc, $\Delta$$Z$=-0.12 kpc.  {\it Left-bottom panel:} projected space
velocity vectors of IC~4665 with respect to the motion of its centre. {\it Right-bottom:}
same as for IC~4665, for Collinder~350.
}
\label{fig6}
\end{figure*}

Figure ~\ref{fig6} illustrates a 3D view of the selected stars, which shows that not only IC~4665 but also Collinder~350 has an elongated shape. In order to build the figure we used the geometrical
middle point between both clusters as the reference framework for positions and space motions,  as this frame is also helpful to highlight their collision.
Therefore, we used Galactic coordinates relative to that midpoint which is
located at ($X$,$Y$,$Z$)= (-7.82,0.17,0.12) kpc, and have a velocity relative to the Galactic 
centre of ($V_X$,$V_Y$,$V_Z$)=(5.54,231.80,5.41) km/s. We performed this Galactic coordinate 
transformation in order to highlight the relative motions of both clusters. Indeed,
the direction of the stellar space velocity vectors reveals that Collinder~350 and IC~4665
are traveling in opposite directions (in this adopted frame of reference). We refer the reader to the analysis of the disrupting open cluster Ruprecht~147, which shows similar stellar structures \citep{yehetal2019}. We also constructed the right panel of Fig.~\ref{fig6} that confirms the  collisional nature of the interaction between these two clusters; where blue/red arrows are computed with 
respect to the assumed centre of mass (middle point between both clusters). By using the centres 
of mass of each clusters and their respective stars, we built Fig~\ref{fig6} (bottom panels), which
illustrate the internal motions of them. As can be seen, stars at opposite sides of the clusters
move in opposite directions. Although the number of stars considered is small, these trends
suggest that the clusters are in the process of disruption.

\begin{figure*}
\includegraphics[width=2\columnwidth]{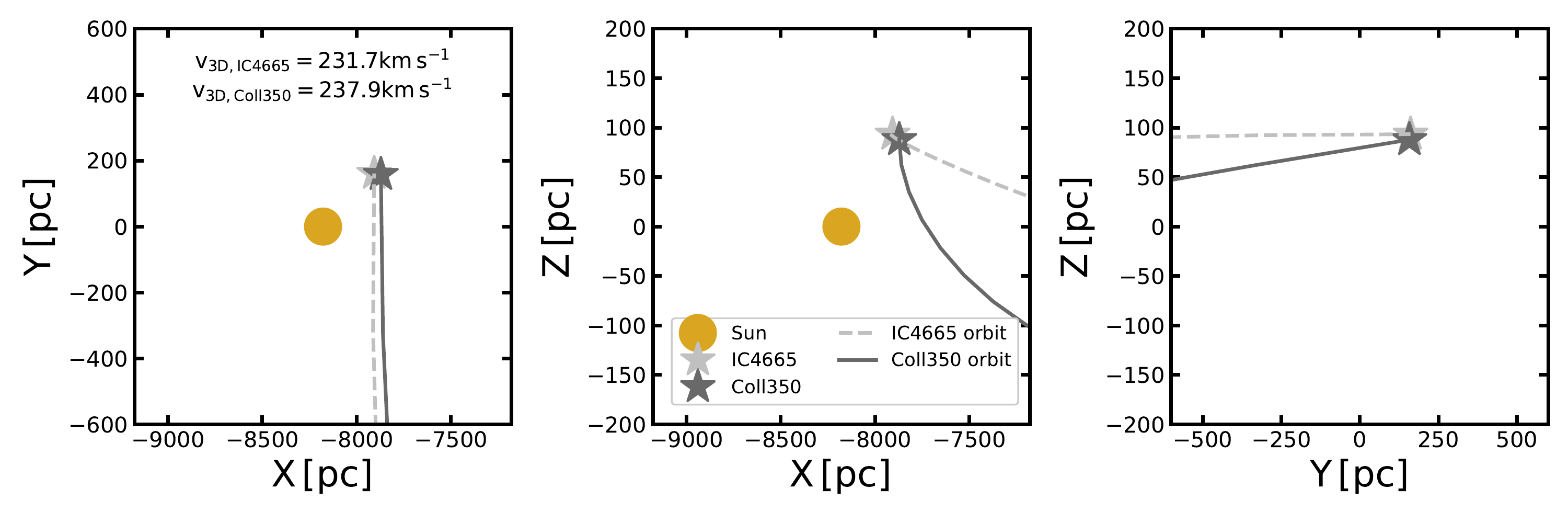}
\caption{The figure is in the Cartesian coordinates, showing the present day positions and past orbits of IC~4665 and Collinder~350. It clearly reveals that these two clusters were physically 
 separated in the past, and it is for the first time that they have come so close to each other at the present day. The total linear Galactocentric velocities of the two clusters are provided in the plot (Collinder~350 is moving $\sim5\kms$ faster that IC~4665).}
\label{fig7}
\end{figure*}
\begin{figure}
\includegraphics[width=\columnwidth]{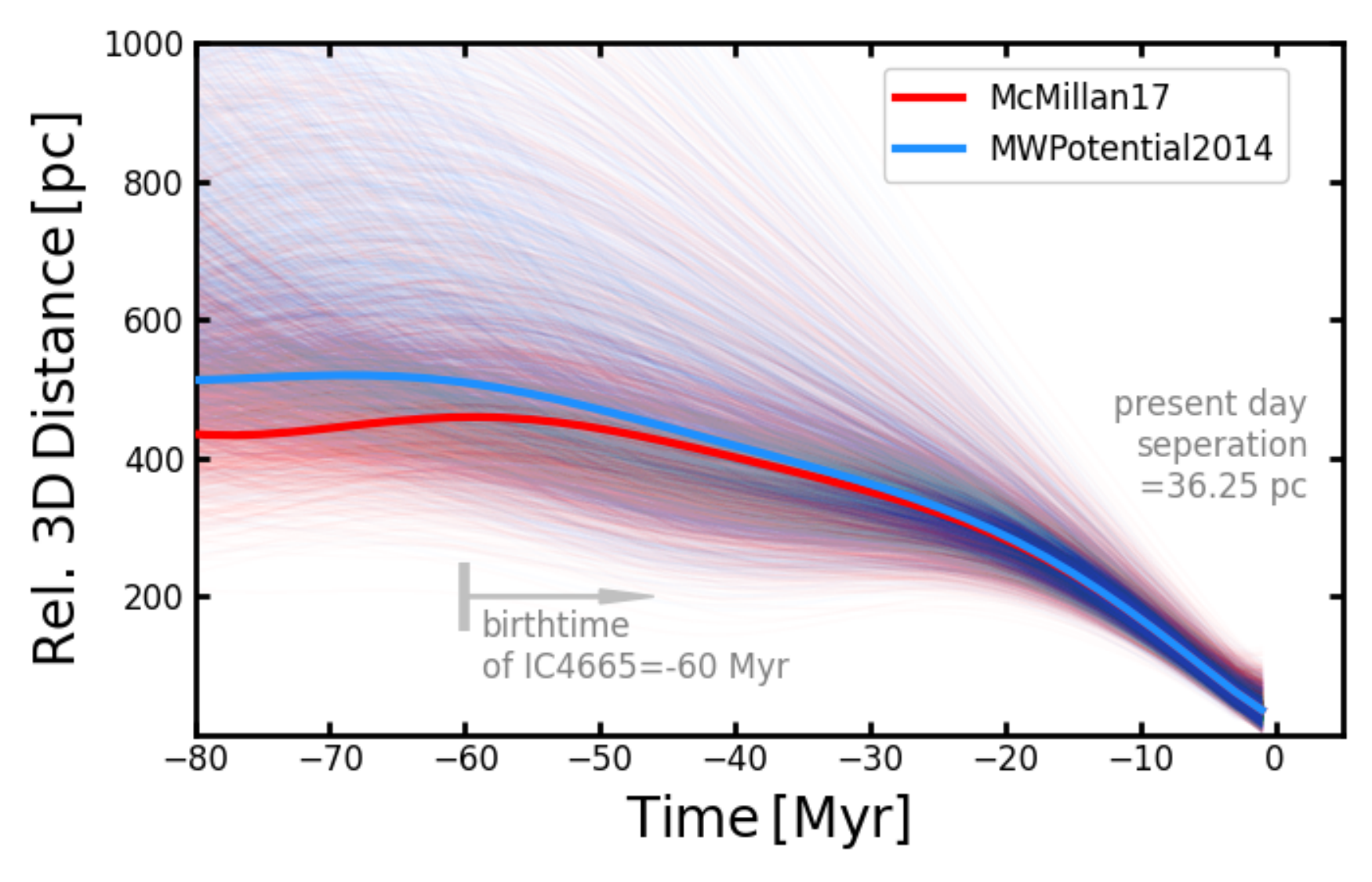}
\caption{Time evolution of the relative 3D distance between IC~4665 and Collinder~350 ($T=0$ represents the present day). We integrated orbits in two Galactic potential models (namely \texttt{MWPotential2014} and \texttt{McMillan17}). The bold curves represent the orbits computed using mean phase-space positions of the clusters. The light shaded curves correspond to the orbits computed by sampling over the measured uncertainties in phase-space values.}
\label{fig8}
\end{figure}
\begin{figure}
\includegraphics[width=\columnwidth]{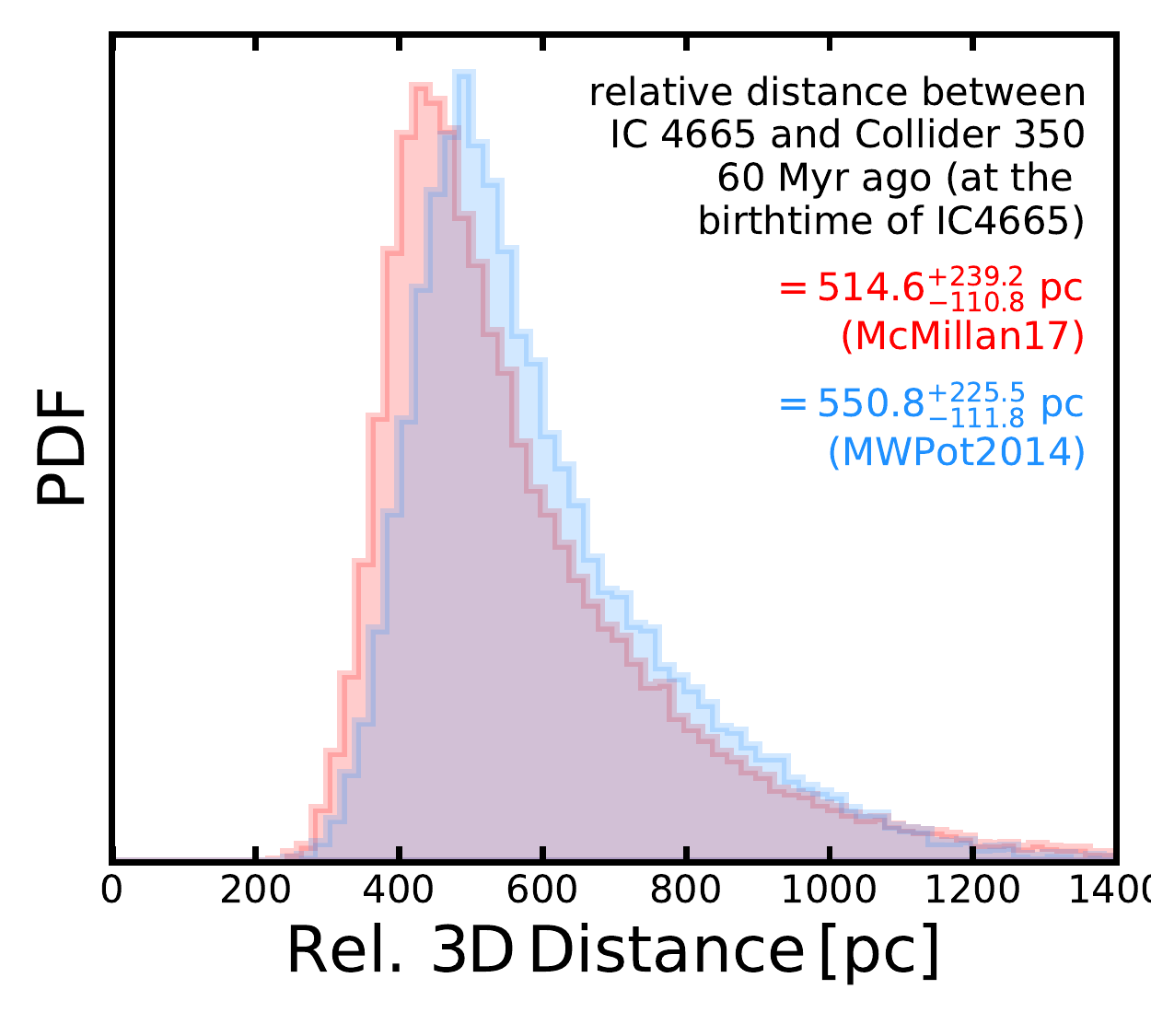}
\caption{The probability distribution function (PDF) of the relative (3D) distance between IC~4665 and Collinder~350 in the past ($60\myr$ ago, at the birth time of IC~4665). This plot is essentially obtained by marginalising over the uncertainties in Figure~\ref{fig6} at $T=-60\myr$. Note that for both the trial Galactic potential models, we obtain a very similar PDF. For instance, the \texttt{McMillan17} model indicates that $60\myr$ ago, the physical 
 separation between the two clusters  was $515^{+239}_{-111}\pc$. This suggests that IC~4665 and Collinder~350 originated at different physical location in the Galaxy, and they represent independent star clusters (and not a binary system).}
\label{fig9}
\end{figure}

\section{The cluster orbits}

It is important to understand whether IC~4665 and Collinder~350 represent the merging of two objects in a binary cluster system (where the two clusters were originally formed together), or whether we are witnessing an actual collision between two independent star clusters. One way to probe this scenario is to integrate the orbits of these two clusters backwards in time and then examine the physical  separation between the two clusters in the past, particularly at the birth time of the younger cluster IC~4665 ($60\myr$ ago). If this physical  separation in the past turns out to be large enough, then that would imply that IC~4665 and Collinder~350 have distinct physical origin, and thus represent independent star clusters. We note also that their ages differ by $\simgt$ 500 Myr.

To compute the orbits of these two open clusters (that are located in the Galactic disk), we follow a similar procedure described in \cite{2021arXiv210409523M} to integrate the orbits of globular clusters (that are otherwise located in the Galactic halo). Briefly, we use the measured heliocentric phase-space positions of the clusters and integrate them backward in time for $\sim80\myr$. To integrate orbits, we try two different Galactic mass models that are amongst the best recommended in the literature -- \texttt{McMillan17} from \cite{2017MNRAS.465...76M} and \texttt{MWPotential2014} from \cite{2015ApJS..216...29B}. Both the models are quite similar in construction: they both are axisymmetric models, each comprising a bulge, disk components and an NFW halo; except that \texttt{McMillan17} overall is slightly heavier than \texttt{MWPotential2014}. The main reason for trying these two different potential models is to ensure that the final results are independent of the assumed Galactic mass model. Finally, in order to account for the measured uncertainties in phase-space associated with the clusters, we sample $1000$ orbits for each cluster. The orbit analysis indicates that both the clusters have very disk-like orbits, with $Z$-component of angular momentum $L_z\sim -1830\kms\kpc$ (implying prograde motion) and eccentricity$\sim 0 - 0.1$ (implying circular trajectories), although Collinder~350 has a slightly larger value of $L_z$ than IC~4665,
due to Collinder~350's larger velocity.

Figure~\ref{fig5} compares the past orbits of IC~4665 and Collinder~350. It immediately reveals that the two clusters were physically  separated in the past, but their orbital evolution has brought them closer to each other at the present day. We infer that the two clusters have only slightly different velocities, with Collinder~350 moving $\sim5\kms$ faster than IC~4665. 

Figure~\ref{fig6} shows the relative (3D) physical  separation between the clusters as a function of the lookback time. The two bold curves correspond to those orbits computed using the mean of the measured phase-space positions of the two clusters. The light shaded curves correspond to the orbits computed by sampling over the phase-space uncertainties. As can be easily discerned, this figure implies that IC~4665 and Collinder~350 were physically  separated in the past, and that it is for the first time that they have come so close to each other at the present day. This implies that the two clusters have very different physical origin and were not born at the same location in our Galaxy. In other words, IC~4665 and Collinder~350 are not binary systems, but independent open clusters. This confirms our 
initial hypothesis that these two clusters represent a scenario of an actual collision between two independent star clusters.

It is also interesting to quantify the physical separation between IC~4665 and Collinder~350 at the birth time of the younger cluster IC~4665. For this, we essentially take a slice of Figure~\ref{fig6} at $T=-60\myr$, and marginalise over the uncertainties to produce Figure~\ref{fig7}. With this we infer that $60\myr$ ago, IC~4665 and Collinder~350 were physically  separated by a distance of $\simgt 500\pc$  (specifically by $515^{+239}_{-111}\pc$ as per \texttt{McMillan17} model and $551^{+226}_{-112}\pc$ as per \texttt{MWPotential2014} model). As one can note, this relative distance between the two clusters is much larger than their combined physical radii. 

\section{Discussion and Conclusion}

We have provided the first evidence of an on-going collision between two star clusters, namely IC~4665 and Collinder~350. This result is based on the analysis of the stellar density mapping of the two clusters, their ages, their present phase-space configuration and also their past orbital evolution in the Galaxy. Since these two clusters were formed at different times ($\Delta$age$\la 500\myr$) and at different locations in the galaxy (with a physical 
 separation of $\simgt500\pc$), this implies that these two clusters do not represent merging 
of two objcets in a binary system, but instead, what we are witnessing is an actual collision 
between two independently formed star clusters.


In future work, it will be interesting to examine whether IC~4665 and Collinder~350 are coalescing into a single star cluster (that will essentially exhibit multiple stellar population with ages$=53\myr$ and $617\myr$), or whether the two clusters are simply passing-by each other (and undergoing only a tidal interaction) but will ultimately evolve as independent clusters. Future N-body simulations will be very useful to analyse this scenario in detail. In case the simulations find that the two clusters are coalescing, it will be important to learn about the final dynamical properties of the resulting system -- does this merged system possesses a net rotation? does the density of this system differs from that of the initial colliding clusters? Moreover, if the simulations indeed favour the ``coalescing'' scenario, then that would further imply that observations of age/metallicity spread in star clusters do not necessarily always require multiple episodes of star formation, but that it can also occur as a result of collision between two star clusters hosting different stellar populations. This would effectively mean that the coalescence of multiple systems may be important in the history of at least some star clusters. Future analysis of IC~4665 and Collinder~350 will allow us to learn more about the detectable observational signatures of collision between independent star clusters.

The on-going collision between IC~4665 and Collinder~350 is not the prototypical phenomenon proposed by theories of the formation of star clusters, however, its discovery is both encouraging and interesting. Therefore, these two clusters provide a unique laboratory to explore various new aspects of formation and evolution theory of star clusters.

\section{Data availability}

Data used in this work are available upon request to the first author.

\section*{Acknowledgements}
We warmly thank Wilton Dias for providing with the tables of membership
probabilities and astrometric information for stars in the field of the studied open clusters, and his catalogue of astrophysical properties of 1743 open clusters.
KM acknowledges support from the Alexander von Humboldt Foundation at Max-Planck-Institut f\"ur Astronomie, Heidelberg. KM is also grateful to the IAU's Gruber Foundation Fellowship Programme for their financial support.
We thank the referee for the thorough reading of the manuscript and
suggestions to improve it.










\bsp	
\label{lastpage}
\end{document}